\documentstyle[prb,aps]{revtex}
\begin{document}

\title{Asymptotically exact mean field theory for the Anderson model
including double occupancy}
\author{J. Holm, R. Kree, and K. Sch\"onhammer}
\address{Institut f\"ur Theoretische Physik, Universit\"at
G\"ottingen,\\
         Bunsenstra{\ss}e 9, D-3400 G\"ottingen, Germany}
\date{March 1993}
\maketitle
\begin{abstract}
The Anderson impurity model for finite values of the Coulomb
repulsion $U$ is studied using a slave boson representation for the
empty and doubly occupied $f$-level. In order to avoid well known
problems with a naive mean field theory for the boson fields, we use
the coherent state path integral representation to first integrate
out the double occupancy slave bosons. The resulting effective action
is linearized using {\bf two-time} auxiliary fields. After integration
over the fermionic degrees of freedom one obtains an effective action
suitable for a $1/N_f$-expansion. Concerning the constraint the same
problem remains as in the infinite $U$ case. For $T \rightarrow 0$
and $N_f \rightarrow \infty$ exact results for the ground state
properties are recovered in the saddle point approximation. Numerical
solutions of the saddle point equations show that even in the
spindegenerate case $N_f = 2$ the results are quite good.
\end{abstract}

\narrowtext
\section{Introduction}
\setcounter{equation}{0}

The Anderson impurity model (AIM) \cite{1} was proposed to gain insight
into the physics of dilute magnetic alloys. Variations of this model
describe e.g. thermodynamic, transport and spectroscopic properties
of Ce mixed valence compounds surprisingly well \cite{2,3}. The strong
on-site $f$-repulsion $U$ makes these compounds typical examples of
strongly correlated electron systems.

Various reviews exist on the theoretical approaches to obtain exact
or approximate solutions for the thermodynamics and correlation
functions of the AIM \cite{3,4,5}. While exact solutions for thermodynamic
properties are known for special cases using the Bethe-Ansatz
\cite{5},
the progress towards controlled approximations for spectral functions
was mainly due to large $N_f$ expansions, where $N_f$ is the
degeneracy of the $f$-level \cite{3,4}.

The two special cases for which the exact thermodynamics can be
obtained via the Bethe-Ansatz are the {\bf infinite $U$ limit} ($U =
\infty, N_f $ arbitrary )
and the {\bf spindegenerate symmetric case} $(2 \varepsilon_f + U = 2
\mu, N_f = 2)$ where $\mu$ is the chemical potential which will be
chosen as the zero of the one electron energies in the following and
$\varepsilon _f$ is the bare $f$-level energy. The one-particle spectral
functions which determine photoemission and inverse photoemission
spectra are rather well understood in the infinite $U$ limit for
arbitrary temperatures and degeneracies down to $N_f = 2$. For
temperatures of the order of the Kondo temperature $T_K$ or smaller
and $f$-occupancies close to one (spinfluctuation limit) there is a
sharp "Kondo-resonance" slightly above the Fermi energy and a broad
resonance near $\varepsilon _f$. The knowledge about the position, the
width and the weight of the Kondo resonance comes from three
different approaches. It was first obtained for zero temperature
using a variational ansatz for the ground state and a $1/N_f$
classification of intermediate states \cite{6,7}. The noncrossing
approximation (NCA) which can be derived by different techniques
works well for all temperatures except $T \ll T_K$ \cite{8,9}. In the
third approach a functional integral (FI) for the partition function
is expressed in terms of an effective action for the slave boson,
which is introduced to handle the constraint of at most single
occupancy of the $f$-level \cite{10,11,12}. If the FI is evaluated in the
saddle point approximation, corresponding to a simple mean field
approximation for the slave boson, the exact ground state energy
\cite{6,7,20} is recovered in the limit $N_f \rightarrow \infty$.
Gaussian fluctuations around the saddle point yield
$(1/N_f)-$corrections in the very low temperature Fermi liquid regime
\cite{11}. Due to the occupancy constraint the saddle point expansion is
not a $(1/N_f)-$expansion in the strict sense. This problem will be
addressed later in this paper for finite values of $U$.

In the spindegenerate symmetric case the Kondo peak is {\bf at} the
Fermi level due to the particle-hole symmetry of the problem
\cite{2,3,4,5}.
For $N_f > 2$ this symmetry is no longer given even for symmetric
parameters $2 \varepsilon _f + U = 0$ and a symmetric band, and the
position of the Kondo peak has not been reliably calculated except in
the large $N_f$ limit \cite{13,14,15}. Also the position of the Kondo peak
for nonsymmetric parameters and $N_f = 2$ is not yet well understood.
For not too low Kondo temperatures the only reliable results are from
quantum Monte Carlo calculations \cite{16}. The spectral function for the
$N_f = 2$-AIM has recently gained renewed attention as its
calculation is required as an intermediate step in the exact solution
of the Hubbard model in infinite dimensions \cite{17,18,19}.

For the finite $U$ AIM and arbitrary degeneracy $N_f$ there exists a
variational approach \cite{13} at $T = 0$ and generalizations of the NCA
\cite{14,15} which go over to the corresponding $U = \infty$ theories
\cite{7},
\cite{8,9} in the large $U$ limit. A naive mean field theory for the slave
bosons for the empty $f$-level as well as the doubly occupied $f$-level
does {\bf not} recover the exact ground state energy in the limit
$N_f \rightarrow \infty$, as it completely misses to obtain the
mathematical structure of an {\bf integral equation} for finite $U$
\cite{13} in contrast to an transcendental equation \cite{6,7,20} for $U =
\infty$. Similar deficiencies of mean field theories for double
occupancy slave bosons occur for the Hubbard model \cite{21,22,23}.

It is the aim of this paper to present a {\bf generalized mean field
theory} for the finite $U$ Anderson model in the $f^0, f^1$ and $f^2$
subspace for arbitrary degeneracy $N_f$, which leads to the {\bf
exact ground state energy in the limit $N_f \rightarrow \infty$}.
Starting from a coherent state path integral we integrate out the
double occupancy bosons, which leads to four - fermion terms. Using a
Stratonovich-Hubbard transformation which introduces {\bf two-time
auxiliary fields} an effective action appropriate for a large $N_f$
expansion is derived, apart from the problem with the occupancy
constraint, similar to the infinite $U$ case \cite{11}. We study
stationary solutions
of the saddle point equations for which the additional auxiliary
fields depend on the {\bf time difference} only. This leads after
Fourier expansion to a system of coupled  nonlinear equations for the
Fourier components. Analytic continuation from the Matsubara
frequencies to the real axis leads to a system of integral equations,
which for finite $N_f$ has to be solved numerically. In the limit
$N_f \rightarrow \infty$ the equations simplify considerably and
for $T \rightarrow 0$ we exactly recover the integral equation
which determines the ground state energy \cite{13}.

In section II we present the model and introduce the coherent state
path integral formulation. An exact effective action for the empty
$f$-level slave boson and the two-time auxiliary fields is derived. In
section III the corresponding saddle point equations are derived. The
simplified equations in the limit $N_f \rightarrow \infty$ are
discussed. The numerical solutions for finite $N_f$ are presented in
section IV. Some results of ref. \cite{13} and details of various
calculations are described in appendices.

\section{The model}

We consider the finite $U$ Anderson impurity Hamiltonian \cite{1,13}
\widetext
\begin{eqnarray} \label{2.1}
H_A & = &
\sum^{N_f}_{\nu=1} \left[ \int \varepsilon c^+_{\varepsilon \nu}
c_{\varepsilon \nu} d\varepsilon  + \varepsilon_f f^+_{\nu} f_{\nu}
+ \int d \varepsilon \left[ V(\varepsilon)f^+_{\nu} c_{\epsilon ,\nu}
+ h.c. \right] \right]
+ U \sum_{{\nu ,\mu} \atop {\nu < \mu}} n_{\nu} n_{\mu} \nonumber\\
& \equiv & H_o + \hat V + \hat U
\end{eqnarray}

\narrowtext
where we have introduced a combined index, $\nu$, for the orbital and
spin degeneracies. Spin-orbit and multiplet splittings are neglected.
The first term describes the conduction states, with energy
$\varepsilon$, and the second term the  $f$-level with the (bare) energy
$\varepsilon _f$. The third term leads to a hopping between these
states where the $V(\varepsilon)$ can  be chosen real and the
last term describes the Coulomb interaction between the $f$-electrons.
The Hamiltonian is studied in the subspace of at most double
occupancy of the $f$-level i.e. in the $f^0, f^1$ and $f^2$ manifold.
This restriction is incorporated using {\bf slave bosons}.

In the infinite $U$ model Coleman \cite{9} has introduced a single slave
boson as a counting device for the number of $f$-electrons. For the
finite $U$ model there are $N_f(N_f-1)/2$ different doubly
occupied $f$-states. Therefore we introduce $N_f(N_f-1)/2$ additional
"heavy bosons" \cite{14}. If $\mid\{0\}>$ denotes the ground state of the
"light" and heavy bosons we make the correspondence

\begin{equation}  \label{2.2}
\mu > \nu : f^+_\mu f^+_\nu \mid\{n_\varepsilon\}>
\longrightarrow h^+_{\mu \nu}\mid\{n_\varepsilon\}>
\otimes \mid\{0\}>
\end{equation}

where $ {\mid\{n_\varepsilon\}>}$ denotes an arbitrary band state and
$h^+_{\mu\nu}$ is the creation operator of a heavy boson.
To remove the restriction $\mu > \nu$ it is useful to define boson
operators for $\mu \leq \nu$ by $h^+_{\mu\nu} := - h^+_{\nu\mu}$ and
$h_{\mu\mu} \equiv 0$.

We now construct a Hamiltonian acting on the product space of
electrons and bosons which is equivalent to H in the $\{ f^0,f^1,f^2\}$
subspace

\widetext
\begin{equation} \label{2.3}
H =  H_0 + \sum_{\mu > \nu}(2\varepsilon_f + U) h^+_{\mu\nu}
h_{\mu\nu}
\quad + \int d\varepsilon \left[\sum^{N_f}_{\nu = 1} V(\varepsilon)
f_\nu^+ c_{\varepsilon\nu}b + \sum^{N_f}_{\mu,\nu = 1}
V(\varepsilon)h^+_{\mu\nu} f_\nu c_{\varepsilon\mu}+ h.c.\right]
\end{equation}
\narrowtext

Here $H_0$ is formally the same as in Eqn. (\ref{2.1}). The Hamiltonian $H$
commutes with the "charge operator" $\hat Q$ defined as

\begin{equation} \label{2.4}
\hat Q \equiv b^+b + \sum_\nu f^+_\nu f_\nu + \sum_{\mu>\nu} h^+_{\mu
\nu}h_{\mu\nu}
\end{equation}

Therefore the subspaces $F_Q$ with definite integer value $Q$ can
be treated separately. In the subspace $F_1$ the Hamiltonian $H$ is
equivalent to the Anderson Hamiltonian $H_A$ in the subspace of
$f^0,f^1$ and $f^2$ states. The partition function is therefore given
by

\begin{equation} \label{2.5}
Z_A = Tr_1 e^{-\beta H}\equiv Z_1
\end{equation}

where the trace is restricted to the subspace of states with $Q = 1$.
This projection to the $Q = 1$ subspace can be expressed as a contour
integral in a complex $\lambda$-plane $(Z \equiv Z_A)$
\begin{equation} \label{2.6}
Z = \frac{\beta}{2 \pi i} \int_C d\lambda e^{\beta\lambda}
Tr \left( e^{- \beta H (\lambda)} \right)
\end{equation}

where $H(\lambda)\equiv H + \lambda \hat Q$ and the contour runs from
$\lambda_R - i\pi/\beta$ to $\lambda_R + i\pi/\beta$, where the real
$\lambda_R$ is arbitrary. Since the trace has now to be performed
over the full Hilbert space of fermions and bosons $Z$ can be expressed
as a coherent state functional integral

\widetext
\begin{eqnarray}\label{2.7}
Z  & = & \int_C \frac{\beta d\lambda}{2 \pi i} e^{\beta\lambda}
\int Df Dc Db Dh
\exp\left\{ -\int^\beta_0 (L_0(\tau)+L_1(\tau))d\tau \right\}\\
\mbox{where} & & \nonumber\\
L_0(\tau) & = & \sum^{N_f}_{\nu = 1}
\left\{\int d\varepsilon \bar{c}_{\varepsilon\nu}(\tau)
\left[\frac{\partial}{\partial\tau} + \varepsilon \right]
c_{\varepsilon\nu}(\tau) +
\bar{f}_\nu(\tau) \left[\frac{\partial}{\partial\tau}+ \varepsilon_f +
\lambda \right]f_\nu(\tau)\right\} \nonumber \\
& &  +b^*(\tau) \left[ \frac{\partial}{\partial\tau} +
\lambda\right]
b(\tau) +
\sum_{\mu > \nu}h^*_{\mu\nu}(\tau) \left[
\frac{\partial}{\partial\tau} + 2 \varepsilon_f + U + \lambda\right]
h_{\mu\nu}(\tau)\\
\mbox{and}& & \nonumber\\
L_1(\tau) & = & \sum^{N_f}_{\nu = 1}
\int d\varepsilon \left[V(\varepsilon) \bar{f}_\nu
(\tau)c_{\varepsilon\nu}(\tau) b(\tau)+
V(\varepsilon) b^*(\tau)\bar{c}_{\varepsilon\nu}(\tau)f_\nu(\tau)
\right]
\nonumber \\
& &  + \sum^{N_f}_{\nu,\mu = 1}
\int d\varepsilon \left[V(\varepsilon)h^*_{\mu\nu}(\tau) f_\nu
(\tau)c_{\varepsilon\mu}(\tau) +
V(\varepsilon) \bar{c}_{\varepsilon\mu}(\tau)\bar{f}_\nu(\tau)
h_{\mu\nu}(\tau) \right]
\end{eqnarray}
\narrowtext

with $\tau$-dependent Grassmann fields $c,\bar{c}$ and $f,\bar{f}$
and complex fields $b$ and $h_{\mu\nu}$.

As the Grassmann fields enter the action $S = \int^{\beta}_0 d\tau
L(\tau)$ with $L(\tau) \equiv L_0(\tau) + L_1(\tau)$ quadratically it
would be possible at this point to eliminate all fermionic variables
by tracing over in the usual way. As we want to construct a mean field
theory which becomes asymptotically exact in the large degeneracy and
small temperature limit this procedure turns out {\bf not} to be
successful. The saddle point equations for the resulting effective
action $S_{eff}(b,h)$ do {\bf not} produce the exact $T = 0$ results
\cite{13} in the large degeneracy limit. This failure can be related to
the fact that the term $L_{1h}(\tau)$ in $L_1(\tau)$ is not well
suited for a large $N_f$ saddle point expansion,because it
consists of $N_f^2$ terms in which different orbital indices $\mu,\nu$
are coupled. To bring this term
in the action into a suitable form, we proceed differently, in a way
which in the first step produces a seemingly more complicated
effective action involving "four-fermion" terms. Factorizing these
terms with a Stratonovich-Hubbard transformation leads to an
interaction term in the effective action of the form needed for a
large $N_f$ expansion.

In the first step the {\bf heavy bosons are integrated out}. The
Gaussian integrals are easily performed and yield using $h_{\nu\mu}
\equiv -h_{\mu\nu}$ in the interaction term $L_{1h} (\tau)$

\widetext
\begin{eqnarray} \label{2.8}
\lefteqn{\int Dh \exp\left\{-\int_0^{\beta}\left[
\sum_{\mu>\nu}h^*_{\mu\nu}(\tau)(\frac{\partial}{\partial\tau}
+2\varepsilon_f +U+\lambda)h_{\mu\nu}(\tau)+L_{1h}(\tau)\right]
\right\}}\nonumber\\
& & =Z^h_0(\lambda)\exp\big\{-\sum_{\mu>\nu}\int d\varepsilon
d\varepsilon'\int_0^{\beta}d\tau \int_0^{\beta}d\tau'
V(\varepsilon)
\left(\bar{c}_{\varepsilon\mu}(\tau)\bar{f}_\nu(\tau)-
\bar{c}_{\varepsilon\nu}(\tau)\bar{f}_\mu(\tau)\right)\nonumber\\
& &\quad G^0_h(\tau-\tau')
\left(f_\nu(\tau')c_{\varepsilon'\mu}(\tau')-
f_\mu(\tau')c_{\varepsilon'\nu}(\tau')\right)V(\varepsilon')\big\},
\end{eqnarray}
\narrowtext

where $Z^h_0(\lambda)=\left( 1-\exp\left[-\beta(2\varepsilon_f
+U+\lambda)\right]\right)^{-N_f(N_f-1)/2}$ is the unperturbed
partition function for the heavy bosons and
\begin{equation} \label{2.9}
G^0_h(\tau-\tau')\equiv
-(\tau\mid(\partial_{\tau}+2\varepsilon_f+U+\lambda)^{-1}\mid \tau')
\end{equation}

is the corresponding unperturbed propagator with the usual bosonic
boundary conditions.

The exponent on the rhs of Eqn. (\ref{2.8}) can be written in a form
suitable for a Stratonovich-Hubbard transformation introducing

\begin{equation} \label{2.10}
Vc_\mu(\tau) \equiv \int d\varepsilon V(\varepsilon)
c_{\mu\varepsilon} (\tau)
\end{equation}

with $V \equiv (\int \mid V(\varepsilon)\mid^2 d\varepsilon)^{1/2}$
and using

\widetext
\begin{eqnarray} \label{2.11}
& & \sum_{\mu > \nu}(\bar c_\mu (\tau) \bar f _\nu (\tau) -
\bar c_\nu (\tau) \bar f _\mu (\tau)) (f_\nu (\tau') c_\mu (\tau')
- f_\mu(\tau')c_\nu (\tau'))  \\
& & = (\sum_\mu \bar c_\mu(\tau) c_\mu (\tau')) (\sum_\nu \bar f _\nu
(\tau) f_\nu (\tau'))
+ (\sum_\mu \bar c_\mu(\tau)f_\mu (\tau')) (\sum_\nu \bar
f_\nu(\tau)c_\nu(\tau')). \nonumber
\end{eqnarray}
\narrowtext

As both terms on the rhs have a simple product form we can use the
identity

\begin{equation} \label{2.12}
e^{CD} = A \int \frac{dRe\alpha \; dIm \alpha}{\pi}
e^{-A\mid\alpha\mid^2 - \sqrt {A} (\alpha C + \alpha^* D)}
\end{equation}

valid for commuting variables $C$ and $D$ and $Re(A) > 0$. The
auxiliary fields which have to be introduced to linearize the
four-Grassmann terms depend on {\bf two time variables}

\widetext
\begin{eqnarray}\label{2.13}
& & \exp \left\{ -V^2 (\sum_\mu \bar c_\mu (\tau) c_\mu (\tau'))
(\sum_\nu \bar f_\nu (\tau) f_\nu(\tau')) G^0_h (\tau-\tau')
\right\}\nonumber \\
& = & \frac{-1}{G^0_h(\tau-\tau')} \int d\mu (F(\tau,\tau'))
\exp\left\{\frac{\mid F(\tau,\tau')\mid ^2}{G^0_h(\tau-\tau')}
- F(\tau,\tau') V \sum_\mu \bar c_\mu (\tau)c_\mu(\tau')
- F^*(\tau,\tau') V \sum_\mu \bar f_\mu (\tau) f_\mu
(\tau') \right\} \nonumber
\end{eqnarray}
\narrowtext

where $d\mu (F) = d ReF dImF/\pi$. The second term in Eqn.
(\ref{2.11})
yields

\widetext
\begin {eqnarray} \label{2.14}
& & \exp \left\{ -V^2 (\sum_\mu \bar c_\mu (\tau)f_\mu(\tau'))
(\sum_\nu \bar f_\nu(\tau)c_\nu(\tau'))G^0_h(\tau
-\tau')\right\}\nonumber \\
& = &  \frac{-1}{G^0_h(\tau -\tau')}\int d\mu (X(\tau,\tau'))
\exp \left\{ \frac{\mid X(\tau,\tau')\mid^2}{G^0_h(\tau-\tau')}
- V \sum_\mu (X(\tau,\tau') \bar c_\mu (\tau) f_\mu (\tau')
+ X^*(\tau,\tau')\bar f_\mu (\tau)c_\mu (\tau'))\right\}\nonumber
\end{eqnarray}
\narrowtext

The linearization produces additional terms bilinear in the Grassmann
variables. The first Grassmann term in the exponent on the rhs of Eqn.
(\ref{2.13}) corresponds to a separable  perturbation $\sim \mid c,\mu ><
c,\mu\mid$, where

\begin{equation} \label{2.15}
\mid c,\mu> := \frac{1}{V} \int d\varepsilon V(\varepsilon)\mid
\varepsilon,\mu>
\end{equation}

are linear combinations of conduction states localized around the
impurity. The other term in the exponent on the rhs of Eqn.
(\ref{2.13})
leads to a fluctuating $f$-level position $\varepsilon_f + \lambda
\rightarrow \varepsilon_f + \lambda + VF^*(\tau,\tau')$. The
Grassmann terms in the exponent on the rhs of Eqn. (2.14) modify the
hopping strength $Vb(\tau)\delta(\tau-\tau') \rightarrow
V(b(\tau)\delta(\tau-\tau')) + X^*(\tau,\tau'))$.

At the next step the Grassmann variables are formally integrated out

\begin{equation} \label{2.16}
Z = \int_c \frac{\beta d \lambda}{2\pi i}
e^{\beta\lambda}Z^h_0(\lambda)
\int DbDFDX e^{-S_{eff}}
\end{equation}

where the effective action $S_{eff}$ is given by

\widetext
\begin{eqnarray} \label{2.17}
S_{eff }& = &\int^\beta_0 b^*(\tau)(\partial_\tau + \lambda)b(\tau)d\tau
- \int^\beta_0 d\tau_1 \int^\beta_0 d\tau_2
\frac{\mid X(\tau_1,\tau_2)\mid^2 + \mid F(\tau_1,\tau_2)\mid^2}
{G^0_h(\tau_1 - \tau_2)}
- N_f Tr \, \ln [\partial \otimes 1_e + 1_\tau \otimes h_0
+ h_1]
\end{eqnarray}
\narrowtext

and the trace runs over the space of antiperiodic functions on the
interval $[0,\beta]$ as well as the space of single electron states
for a single channel. The operators $h_0$ and $h_1$ are given by

\begin{eqnarray} \label{2.18a}
h_0 & = & \int \varepsilon \mid \varepsilon > <\varepsilon \mid
d \varepsilon + (\varepsilon_f + \lambda) \mid f><f \mid\\
h_1 & = & V(F^*\mid f><f\mid + F\mid c><c \mid
\label{2.18b} \label{2.18last}\\
& & \, + (b^*+X)\mid c><f \mid + (b + X^*)\mid f><c \mid) \nonumber
\end{eqnarray}

The label of the electronic channel has been supressed as each
channel gives the same contribution leading to the factor $N_f$ in front
of the trace. Scaling all Bose fields by a factor $\sqrt {N_f}$ i.e.
$b \rightarrow \sqrt{N_f} B$ etc. the effective action becomes
proportional to $N_f$ and a well defined large $N_f$ expansion could
be carried out, if not the same problem known from the infinite $U$
case \cite{11} due to the constraint integral over $\lambda$ would show
up. The exponent $\beta\lambda$ is {\bf not} proportional to $N_f$
which leads to a similar behaviour as discussed in detail for $U =
\infty$ \cite{11}.

Nevertheless we treat the functional integral in Eqn. (\ref{2.16}) by a
saddle point approximation (without scaling of the Bose fields).

\section{The saddle point equations}

To derive the saddle point equations (SPE) the Bose fields and their
complex conjugates $(F,F^*\mbox{etc.})$ are treated as independent
variables and we replace $F^* \rightarrow \bar F$ etc. to
indicate that the saddle point values of $F$ and $\bar F$ etc. are in
general not complex conjugate fields.

The functional derivatives of the effective action $S_{eff}$ involve
single electron propagators which we abbreviate as $(i,j=f\mbox{ or }
c)$
\widetext
\begin{equation} \label{3.1}
G_{ij} (\tau,\tau') \equiv - (\tau \mid <i\mid [\partial \otimes 1_e
+ 1_\tau \otimes h_0+h_1]^{-1}\mid j>\mid\tau')
\end{equation}
\narrowtext
which obey the usual fermionic boundary conditions. As the factor
$e^{\beta\lambda} Z_0^h (\lambda)$ is independent of the Bose fields
the SPE from the derivatives with respect to the Bose fields read
\begin{eqnarray}\label{3.2a}
0 & = & \frac{\delta S_{eff}}{\delta \bar b(\tau)}
= (\partial_\tau + \lambda) b(\tau) + N_f V G_{fc} (\tau,\tau)\\
0 & = & \frac{\delta S_{eff}}{\delta b(\tau)}
= (-\partial_\tau + \lambda) \bar b(\tau) + N_f V G_{cf} (\tau,\tau)
\\
0 & = & \frac{\delta S_{eff}}{\delta X(\tau,\tau')}
= -\frac{\bar X (\tau,\tau')} {G^0_h(\tau - \tau')} + N_f V G_{fc}
(\tau',\tau) \\
0 & = & \frac{\delta S_{eff}}{\delta \bar X(\tau,\tau')}
= -\frac{X (\tau,\tau')} {G^0_h(\tau - \tau')} + N_f V
G_{cf}(\tau',\tau) \\
0 & = & \frac{\delta S_{eff}}{\delta F(\tau,\tau')}
= -\frac{\bar F (\tau,\tau')} {G^0_h(\tau - \tau')} + N_f V
G_{cc}(\tau',\tau)\\
0 & = & \frac{\delta S_{eff}}{\delta \bar F(\tau,\tau')}
= -\frac{F(\tau,\tau')} {G^0_h(\tau - \tau')} + N_f V
G_{ff}(\tau'\tau)\label{3.2last}
\end{eqnarray}

The generalization of the static approximation in the $U = \infty$
case amounts to search for solutions with "time"-translational
invariance, i.e. the two-time Bose fields depend on the
"time"-difference only, while $b$ and $\bar b$ are constant. This
yields the equations
\begin{eqnarray} \label{3.3a}
-\lambda b & = & N_f\,V\,G_{fc}(0)\\
-\lambda \bar b & = & N_f\,V\,G_{cf}(0)  \label{3.3b}\\
\bar X (\tau - \tau') & = & N_f\,V\,G^0_h(\tau - \tau') G_{fc}(\tau'
- \tau)  \label{3.3c}\\
X (\tau - \tau') & = & N_f\,V\,G^0_h(\tau - \tau') G_{cf}(\tau'
- \tau)  \label{3.3d}\\
\bar F (\tau - \tau') & = & N_f\,V\,G^0_h(\tau - \tau') G_{cc}(\tau'
- \tau)  \label{3.3e}\\
F (\tau - \tau') & = & N_f\,V\,G^0_h(\tau - \tau') G_{ff}(\tau'
- \tau) \label{3.3f}\label{3.3last}
\end{eqnarray}

In the SPE from the constraint variable $\lambda$ we have also to
take into account the $\lambda$-dependence of the integrand in
(\ref{2.13})
not contained in $S_{eff}$. As we are interested in the low
temperature regime the partion function $Z^h_0(\lambda)$ can be
approximated by 1, but the factor $e^{\beta\lambda}$ has to be kept.
This yields
\widetext
\begin{eqnarray} \label{3.4}
0 & = & \frac{\delta(S_{eff} - \beta\lambda)}{\delta\lambda} \\
& = & -\int \bar b(\tau_1) b(\tau_1) d\tau - N_f \int G_{ff}(\tau,\tau +
0)d \tau - \beta \nonumber \\
& & \;
+ \int (X(\tau_1,\tau_2)\bar X (\tau_1,\tau_2) +
(\bar F(\tau_1,\tau_2)\bar
F(\tau_1,\tau_2))\frac{\partial}{\partial\lambda}(1/G^0_h(\tau_1-\tau_
2))d\tau_1 d\tau_2 \nonumber
\end{eqnarray}
\narrowtext

Using Eqn. (\ref{3.3c}-\ref{3.3last}) this SPE reads $((n_f)_{ps} \equiv N_f
G_{ff}(-0))$
\begin{equation} \label{3.5}
\mid b \mid^2+(n_f)_{ps}+A =1
\end{equation}
where $A$ is given by $(\tilde{V}^2 \equiv N_fV^2)$
\widetext
\begin{equation} \label{3.6}
A = \tilde{V}^2 \int d\tau [N_f G_{fc} (-\tau) G_{cf} (-\tau)
+ N_f G_{ff}(-\tau) G_{cc} (-\tau)]
\frac{\partial}{\partial\lambda} G^0_h(\tau)
\end{equation}
\narrowtext

As will be shown in appendix C, A presents the probability for the
double occupancy of the $f$-level. To solve the SPE
(\ref{3.3a}-\ref{3.3last}) and
(\ref{3.5})
we write the propagators as Fourier series
\begin{eqnarray}\label{3.7a}
G^0_h (\tau-\tau')
& = &
\frac{1}{\beta}\sum_m e^{-i\nu _m (\tau-\tau')}
G^0_h(i\nu_m) \\
G_{ij}(\tau' - \tau)
& = &
\frac{1}{\beta}\sum_h e^{-i\omega_n (\tau'-\tau)}
G_{ij}(i\omega_n)\label{3.7last}
\end{eqnarray}

where the Matsubara frequencies $\nu_m = 2\pi m/\beta \hbar,
m \in \mbox{\bf Z}$ are of the Bose type, while the $\omega_n = 2
\pi(n+1)/\beta \hbar, n \in \mbox{\bf Z}$ are fermionic frequencies. Then
Eqns. (\ref{3.3a}-\ref{3.3last}) lead to equations for the Fourier
coefficients, e.g.

\begin{equation} \label{3.8}
\bar X (i\omega_n) = N_fV (\frac{1}{\beta} \sum_m G^0_h (i\nu_m)
G_{fc}(i\nu_m - i\omega_n))
\end{equation}

and corresponding equations for the other Bose fields. To calculate
the $G_{ij}(i\omega_n)$ one has to calculate the resolvent of the one
electron operator $h(i\omega_n) \equiv h_0 + h_1(i\omega_n)$

\begin{equation} \label{3.9}
G_{ij}(i\omega_n) = <i\mid (i\omega_n - h(i\omega_n))^{-1} \mid j>
\end{equation}

The inversion is straightforward using a partioning technique with
the projector $P = \mid f>< f\mid$. With
\begin{equation} \label{3.10}
G^{00}_{cc}(z)
\equiv
<c\mid (z-h^M_0)^{-1}\mid c>
\end{equation}
and
\begin{equation}\label{3.11}
\quad G^0_{cc}(i\omega_n)
\equiv
G^{00}_{cc}(i\omega_n)/(1-VF(i\omega_n)G^{00}_{cc}(i\omega_n))
\end{equation}
the result for the f-propagator is given by $(\tilde{\varepsilon}_f
\equiv \varepsilon_f + \lambda)$

\widetext
\begin{eqnarray}\label{3.12a}
G_{ff}(i\omega_n)  =
[i\omega_n - \tilde{\varepsilon}_f - V \bar F(i\omega_n)
 - V^2 (b + \bar X (i\omega_n)) G^0_{cc}(i\omega_n)
(\bar b + X(i\omega_n))]^{-1} \quad
\end{eqnarray}
\narrowtext
The other resolvent matrix elements are
\begin{eqnarray}\label{3.12b}
G_{cf}(i\omega_n) & = &
G^0_{cc}(i\omega_n)V(\bar b + X(i\omega_n))G_{ff}(i\omega_n), \\
G_{fc}(i\omega_n) & = &
G^0_{ff}(i\omega_n)V(b + \bar X(i\omega_n))G^0_{cc}(i\omega_n),
\label{3.12c}\\
G_{cc}(i\omega_n) & = &
G^0_{cc}(i\omega_n) + G_{cf}(i\omega_n) V (b + \bar
X(i\omega_n))\label{3.12d}\label{3.12last}
\end{eqnarray}

Alternatively $G_{cc}$ can be written in the form
\begin{equation} \label{3.13}
G_{cc}(i\omega_n) = (i\omega_n - \tilde{\varepsilon}_f - V \bar
F(i\omega_n)) G_{ff}(i\omega_n) G^0_{cc}(i\omega_n)
\end{equation}

In the following we study the solution of the SPE with $b,\bar b$
different from zero and define
\begin{eqnarray}\label{3.14a}
\bar x(i\omega_n) & \equiv & \bar X(i\omega_n)/b \\
x(i\omega_n) & \equiv &  X(i\omega_n)/\bar b
\end{eqnarray}\label{3.14last}

Then the SPE (\ref{3.3a}-\ref{3.3last}) take the form
\widetext
\begin{eqnarray}\label{3.15a}
- \lambda & = & \tilde{V}^2 \frac{1}{\beta}\sum_n G_{ff}(i\omega_n)
G^0_{cc}(i\omega_n)(1 + \bar x (i\omega_n))\\
\bar x(i\omega_n) & = &
\tilde{V}^2 \frac{1}{\beta}\sum_m G^0_h(i\nu_m)
G_{ff}(i\nu_m - i\omega_n)G^0_{cc}(i\nu_m - i\omega_n)
\qquad(1 + \bar x (i\nu_m - i\omega_n)) \label{3.15b}\\
V\bar F(i\omega_n) & = & \tilde{V}^2 \frac{1}{\beta} \sum_m
G^0_{h}(i\nu_m)G^0_{cc}(i\nu_m - i\omega_n)
\quad(i\nu_m - i\omega_n -
\tilde{\varepsilon}_f - V\bar F (i\nu_m - i\omega_n))
G_{ff}(i\nu_m - i\omega_n)  \label{3.15c}\\
VF(i\omega_n) & = & \tilde{V}^2 \frac{1}{\beta} \sum_m
G^0_{h}(i\nu_m)G_{ff}(i\nu_m - i\omega_n)\label{3.15d}\label{3.15last}
\end{eqnarray}
\narrowtext

The equations following from (\ref{3.3b}) and (\ref{3.3d}) involve
$x(i\omega_n)$. Comparing with Eqns. (\ref{3.15a}) and (\ref{3.15b}) leads to
the
identity $\bar x(i\omega_n) = x(i\omega_n)$, i.e. we can drop the
additional equations. Together with (\ref{3.5}) the Eqns.
(\ref{3.15a}-\ref{3.15last}) determine
the Bose fields at the saddle point.

In order to obtain a first understanding of these equations we start
with a discussion of the limit $N_f \rightarrow \infty$. It is
easy to see that it is consistent to assume that $\bar x(i\omega_n),
VF (i\omega_n)$ and $V\bar F(i\omega_n)$ are of order unity in the
limit $N_f \rightarrow \infty$ with $\tilde {V}$ constant. This implies

\begin{equation} \label{3.16}
G_{ff}(i\omega_n)\stackrel{N_f\rightarrow \infty}{\longrightarrow}
(i\omega_n - \tilde{\varepsilon}_f - V\bar F(i\omega_n))^{-1}
\equiv G^0_{ff}(i\omega_n)
\end{equation}

and $G_{cc}(i\omega_n) \rightarrow G^0_{cc}(i\omega_n)$. This
simplifies Eqns. (\ref{3.15d}) and (\ref{3.15c}) for $F(i\omega_n)$ and $\bar
F(i\omega_n)$

\widetext
\begin{eqnarray}\label{3.17a}
VF(i\omega_n)& = &\tilde{V}^2 \frac{1}{\beta} \sum_m
\frac{1}{[i\nu_m - (2\varepsilon_f + U + \lambda)]}
\cdot
\frac{1}{[i\nu_m - i\omega_n - \tilde{\varepsilon}_f-V\bar F(i\nu_m -
i\omega_n)]}\nonumber \\ \\
V\bar F(i\omega_n)& = &\tilde{V}^2 \frac{1}{\beta} \sum_m
\frac{1}{[i\nu_m - (2\varepsilon_f + U + \lambda)]}
\cdot
\frac{1}{[(G^{00}_{cc}(i\nu_m -i\omega_n))^{-1} - VF(i\nu_m -
i\omega_n)]}\nonumber \\ \label{3.17b}
\end{eqnarray}
\narrowtext

which now constitute a closed system of equations for a given value of
$\lambda$. At this point it is necessary to discuss the strategy to
solve the SPE for the Fourier components. Guided by results of an
expansion in powers of $1/U$ we will assume that an {\bf analytic
continuation} $F(i\omega_n)\rightarrow F(z)$ etc. is possible, with
$F(z)$ etc. {\bf analytic functions except at (parts of) the real
axis}. As $F$ and $\bar F$ are of order $1/U$ the leading order
result for $VF(i\omega_n)$ is obtained by neglecting $V\bar F(i\nu_m
- i\omega_n)$ in the denominator in Eqn. (\ref{3.17a}). Then the Matsubara
sum can be performed in the usual way by contour integration and one
obtains
\begin{equation} \label{3.18}
VF_{0}(i\omega_n) =
\tilde{V}^2 \frac{n_- (2\varepsilon_f + U + \lambda)
+ f(\varepsilon_f + \lambda)}
{i\omega_n - \varepsilon_f - U}
\end{equation}

where $n_-(\varepsilon)$ is the Bose function and $f(\varepsilon)$ is
the Fermi function. The leading order result for $V \bar F(i\omega_n)$
is
\begin{equation} \label{3.19}
V\bar F_{0}(i\omega_n) =
\int \frac{\mid\tilde{V}(\varepsilon)\mid^2
(n_-(2\varepsilon_f + U + \lambda) + f(\varepsilon))}
{i\omega_n + \varepsilon - (2\varepsilon_f + U + \lambda)}
d\varepsilon
\end{equation}

To this order the analytical continuation is trivial and has the
properties discussed above. In the low temperature limit the Bose
functions at $\tilde{\varepsilon}_U \equiv 2\varepsilon_f + U +
\lambda$ can be neglected. At $T = 0$ the function $V\bar F_{(0)}(z)$
has a branch cut from $\tilde{\varepsilon}_U$ to $\tilde{\varepsilon}_U
+ B$. For the solution of Eqns. (\ref{3.17a}, \ref{3.17b}) for arbitrary $U$ we
now {\bf
assume} the spectral representations $(i = c,f)$

\begin{equation} \label{3.20}
G^{(0)}_{ii}(i\omega_n) = \int \frac{\rho_{ii}^{(0)}(\varepsilon)}
{i\omega_n - \varepsilon} d\varepsilon
\end{equation}

Performing the Matsubara sums in Eqn. (\ref{3.17a}, \ref{3.17b}) yields after
analytic
continuation

\begin{eqnarray}\label{3.21a}
VF(z) & = & \tilde V^2 \int
\frac{\rho^0_{ff}(\varepsilon)(f(\varepsilon) +
n_-(\tilde\varepsilon_U))}
{z + \varepsilon - \tilde\varepsilon_U}d\varepsilon  \\
V\bar F(z) & = & \tilde V^2 \int
\frac{\rho^0_{cc}(\varepsilon)(f(\varepsilon) +
n_-(\tilde\varepsilon_U))}
{z + \varepsilon - \tilde\varepsilon_U}d\varepsilon  \label{3.21b}
\end{eqnarray}

Together with Eqn. (\ref{3.16}) and $G^0_{cc}(z)^{-1} = G^{00}_{cc}(z)^{-1}
- VF(z)$ this constitutes a system of integral equations for the
spectral weight functions $\rho^0_{ff}$ and $\rho^0_{cc}$. As the
finite temperature theory will be discussed later for arbitrary $N_f$
we restrict ourselves here to the limit $T \rightarrow 0$. At the saddle point
value of $\lambda$ one expects $\rho^0_{ff}(\varepsilon)$ to vanish
for $\varepsilon < \varepsilon_F$ as in the $U = \infty$ case
\cite{10}.
In addition to the Kondo peak slightly above the Fermi energy
$\varepsilon_F = 0$, $\rho^{(0)}_{ff}(\varepsilon)$ has an $f^2$
contribution above $\tilde{\varepsilon}_U$, as can be inferred form
Eqn. (\ref{3.19}). Therefore the integrand on the rhs of Eqn.
(\ref{3.21a})
vanishes, i.e. $VF(z) \equiv 0$ at $T = 0$ in the limit $N_f
\rightarrow \infty$. This leads to $G^0_{cc}(z) = G^{00}_{cc}(z)$ and

\begin{equation} \label{3.22}
G^0_{ff}(z) = (z - \tilde\varepsilon_f + \tilde\Gamma
(2\varepsilon_f + U + \lambda - z))^{-1}
\end{equation}
with $\tilde \Gamma(z)$ defined in (A.6).

For $T \rightarrow 0$ and $N_f \rightarrow \infty$ Eqns. (\ref{3.15a}) and
(\ref{3.15b}) then simplify to

\widetext
\begin{eqnarray}\label{3.23a}
-\lambda  =  \int^{B'}_{-B}\mid \tilde V (\varepsilon)\mid^2
\frac{1}{\beta}\sum_n
& & \frac{1}{[i\omega_n -\tilde\varepsilon_f + \tilde\Gamma (\tilde
\varepsilon_U - i\omega_n)]}
\frac{1 + \bar x(i\omega_n)}
{(i\omega_n - \varepsilon)} d\varepsilon \\
\bar x(i\omega_n)  =  \int^{B'}_{-B} \mid \tilde V (\varepsilon)\mid^2
\frac{1}{\beta}\sum_m
& &\frac{1}{(i\nu_m-\tilde\varepsilon_U)}
\frac{1}{[i\nu_n - i\omega_n -\tilde\varepsilon_f + \tilde\Gamma
(\tilde\varepsilon_U - i\nu_m + i\omega_n)]}
\frac{1 + \bar x (i\nu_m - i\omega_n)}
{(i\nu_m - i\omega_n - \varepsilon)} d\varepsilon \label{3.23b}
\end{eqnarray}
\narrowtext

In order to perform the Matsubara sums by contour integration it is
necessary to analytically continue the Fourier  coefficients $\bar
x(i\omega_n)$. A useful guide is again to first calculate $\bar
x(i\omega_n)$ to leading order in $1/U$, which amounts to iterate Eqn.
(\ref{3.23b}). This calculation shows that $\bar x(i\omega_n)$ can be
analytically continued and at $T = 0$ has a branch cut from
$\tilde{\varepsilon}_U$ to $\tilde{\varepsilon}_U+B$ . To evaluate
the Matsubara sums in (\ref{3.23a}) we therefore assume that $\bar x(z)$ has
a spectral representation with a branch cut starting at a positive
energy value. Then for $T \rightarrow 0$ only the last pole term on
the rhs of (\ref{3.23a}, \ref{3.23b}) contributes to the contour integral and
one
obtains after analytic continuation

\widetext
\begin{eqnarray}\label{3.24a}
-\lambda  =  \int^0_{-B}
& & \frac{\mid V(\varepsilon)\mid^2 (1+\bar x(\varepsilon))}
{\varepsilon - \varepsilon_f-\lambda + \tilde\Gamma(2\varepsilon_f +
U+\lambda-\varepsilon)}d\varepsilon\\
\bar x(z) = - \int^0_{-B}
& & \frac{\mid V(\varepsilon)\mid^2 (1+\bar x(\varepsilon))}
{(2\varepsilon_f + U+\lambda -\varepsilon - z)
[\varepsilon - \varepsilon_f-\lambda + \tilde\Gamma(2\varepsilon_f +
U+\lambda-\varepsilon)]}d\varepsilon  \nonumber \\ \label{3.24last}
\end{eqnarray}
\narrowtext

If $z$ is chosen on the real axis $-B < Rez < 0$ the second equation
is an integral equation for $\bar x(\varepsilon)$ with $\varepsilon
\in [-B,0]$ which together with (\ref{3.24a}) determines $\lambda$. With
the identification $\bar x(\varepsilon) \leftrightarrow
c(\varepsilon)$ and $\lambda \leftrightarrow -\Delta E$ these are
exactly the equations which occur in the description of the ground
state in the limit $N_f\rightarrow\infty$ \cite{13}. This is discussed
in appendix A. A direct proof for the relation $-\lambda = \Delta E$
valid at $T=0$ in the limit $N_f\rightarrow\infty$ is presented in
appendix B.

After this discussion of the limit $N_f\rightarrow\infty$ we return to
the full SPE (\ref{3.15a}-\ref{3.15last}) for finite  values  of $N_f$. We use
the
same strategy as for the solution of Equs. (\ref{3.17a},\ref{3.17b}) and
\underline{assume}
that the Greens functions $\tilde{G}_{ij}(i\omega_n)$ can be
analytically continued, and as a function of the complex variable $z$
are analytic expect at (parts of) the real axis. This implies the spectral
representation
\begin{equation} \label{3.25}
   \tilde{G}_{ij}(z) = \int_{-\infty}^{\infty}
\frac{\rho_{ij}(\varepsilon)}{z-\varepsilon} d\varepsilon
\end{equation}

Using these spectral representations the Matsubara sums in Equs.
(\ref{3.15a}-\ref{3.15last}) can easily be performed. If we also introduce the
functions $R_{ij}(z)$,

\begin{equation} \label{3.26}
R_{ij}(z) = \int_{-\infty}^{\infty}
\frac{\rho_{ij}(\varepsilon)f(\varepsilon)}{z-\varepsilon} d\varepsilon
\end{equation}

\noindent
we obtain
\widetext
\begin{eqnarray}\label{3.27a}
VF(z) &=& -\tilde{V}^2 \lbrack
n_-(\tilde{\varepsilon}_U)G_{ff}(\tilde{\varepsilon}_U-z)
+R_{ff}(\tilde{\varepsilon}_U-z)\rbrack  \\
V\bar{F}(z) &=& -\tilde{V}^2 \lbrack
n_-(\tilde{\varepsilon}_U)G_{cc}(\tilde{\varepsilon}_U-z)
+R_{cc}(\tilde{\varepsilon}_U-z)\rbrack  \label{3.27b}\\
X(z) &=& -N_fV \lbrack
n_-(\tilde{\varepsilon}_U)G_{cf}(\tilde{\varepsilon}_U-z)
+R_{cf}(\tilde{\varepsilon}_U-z)\rbrack \label{3.27c} \label{3.27last}
\end{eqnarray}
\narrowtext
Therefore $F$, $\bar{F}$ and $X$ are also analytic except at the real
axis and can also be expressed in terms spectral functions
$\rho_F(\varepsilon)$,
$\rho_{\bar{F}}(\varepsilon)$ and $\rho_{X}(\varepsilon)$. Together with the
relations (\ref{3.12a}-\ref{3.12last}) the equations
(\ref{3.27a}-\ref{3.27last}) provide a set of nonlinear
integral equations for these spectral functions for given values of $\lambda$
and $b$. In order to determine the saddle point values of $\lambda$ and $b$ we
in addition have to use equation (\ref{3.15a}) and the constraint relations
(\ref{3.5}) which was irrelevant in the limit $N_f\rightarrow\infty$. Using the
spectral representation (\ref{3.25}) the probability for double occupancy $A$
(\ref{3.6}) can be expressed in terms of the spectral functions. For
temperatures $k_BT\ll\tilde{\varepsilon}_U$ one obtains

\widetext
\begin{equation} \label{3.28}
A=(N_fV)^2 \int d\varepsilon d\varepsilon'
\frac{\rho_{ff}(\varepsilon)\rho_{cc}(\varepsilon')
      +\rho_{fc}(\varepsilon)\rho_{cf}(\varepsilon')}
{(\tilde{\varepsilon}_U-\varepsilon-\varepsilon')^2} f(\varepsilon)
f(\varepsilon')
\end{equation}
\narrowtext
The numerical solutions of the coupled system of equations is described in the
next section.

\section{Numerical solution of the saddle point equations}

In this section we present an outline of our numerical procedure to solve the
SPE for a
given band density of states, which determines the unperturbed band propagator
$G_{cc}^{00}$. The equations (\ref{3.27a}-\ref{3.27last}) provide relations
between
the spectral functions $\rho_K(\varepsilon)$ with $K\in\{F, \bar{F},
X\}$
and the spectral functions $\rho_{ij}(\varepsilon)$ with
$i,j\in\{f,c\}$.
For $k_BT\ll\tilde{\varepsilon}_U$ they read

\begin{eqnarray} \label{4.1a}
\rho_F(\varepsilon) &= N_fV \rho_{ff}(\tilde{\varepsilon}_U-\varepsilon)
f(\tilde{\varepsilon}_U-\varepsilon)\\
\rho_{\bar{F}}(\varepsilon) &= N_fV
\rho_{cc}(\tilde{\varepsilon}_U-\varepsilon)
f(\tilde{\varepsilon}_U-\varepsilon)\\
\rho_X(\varepsilon) &= -N_fV \rho_{fc}(\tilde{\varepsilon}_U-\varepsilon)
f(\tilde{\varepsilon}_U-\varepsilon)\label{4.1last}
\end{eqnarray}

We solve the SPE iterativly using e.g. the results for
$N_f\rightarrow\infty$ discussed in section III as the starting values
for the $\rho_K$, $\lambda$ and $b$. From the $\rho_K$ we calculate
the functions $K(\varepsilon +i0)$, which via equations
(\ref{3.12a}-\ref{3.12last})
determine the propagators $G_{ij}(\varepsilon +i0)$ and their spectral
functions $\rho_{ij}(\varepsilon)$. Using equations
(\ref{4.1a}-\ref{4.1last}) we
obtain new spectral functions $\rho_K(\varepsilon)$. The constraint
relation (\ref{3.5}) involves the pseudo $f$-occoupancy $(n_f)_{ps}$

\begin{equation} \label{4.2}
(n_f)_{ps} = N_f \int \rho_{ff}(\varepsilon) f(\varepsilon) d\varepsilon
\end{equation}

and the probability for double accupancy $A$, which we evaluate using
(\ref{3.28}) and (\ref{4.1a}-\ref{4.1last})
\begin{equation} \label{4.3}
A=\int d\varepsilon d\varepsilon'
\frac{\rho_{F}(\varepsilon)\rho_{\bar{F}}(\varepsilon')
      +\rho_{X}(\varepsilon)\rho_{X}(\varepsilon')}
{(\tilde{\varepsilon}_U-\varepsilon-\varepsilon')^2} f(\varepsilon)
f(\varepsilon')
\end{equation}

In order to obtain a numerically stable method to determine new
values of $\lambda$ and $b$ we treat equations (\ref{3.5}) and
(\ref{3.3a})
\begin{eqnarray}
1 - (n_f)_{ps} -A -b^2 &= &0 \label{4.4}\\
\lambda b + N_fV\int\rho_{fc}(\varepsilon)f(\varepsilon)d\varepsilon &= &0
\label{4.5}
\end{eqnarray}

as a two dimensional system of equations and solve it by Newton's
method.  We then go to the beginning of the loop and terminate the
iterative procedure when the relative change of all propagators is
less then $10^{-4}$ and $\lambda$ and $b$ solve (\ref{4.4}) and
(\ref{4.5}) with an error less then $10^{-3}$. For the functions
$\rho_K(\varepsilon)$ we usually used a 500 points logarithmic mesh,
which was implemented by logarithmic submeshes to resolve
characteristic features such as band edges.

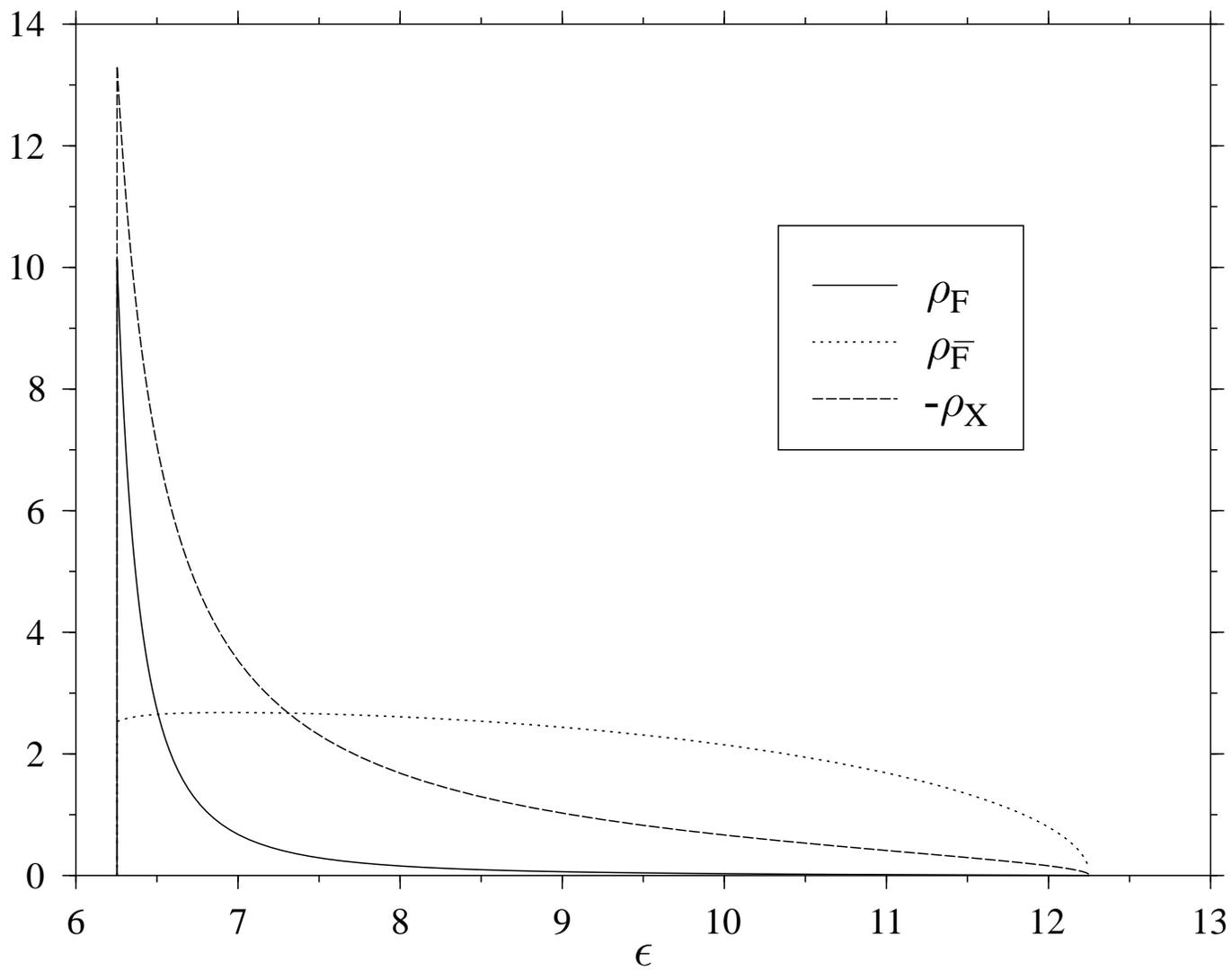
\begin{figure}
\caption{The spectral functions $\rho_F$,
$\rho_{\bar{F}}$ and $\rho_X$ for $N_f=14$, $\varepsilon_f=-2.5$, $U=8$
and $\tilde{\Delta}=2.67$.}
\label{fig1}
\end{figure}

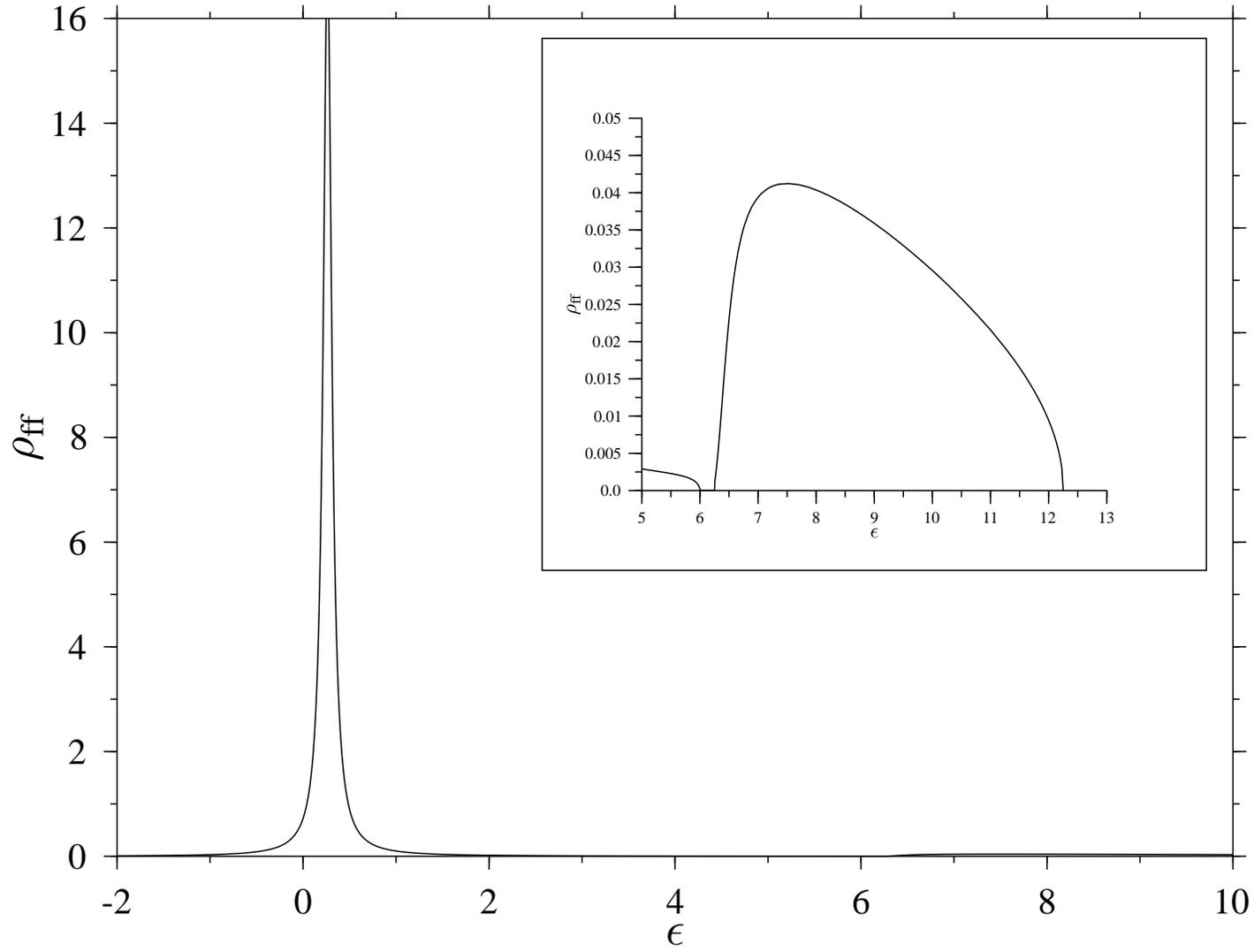
\begin{figure}
\caption{The pseudo $f$-spectral function $(\rho_{ff})_{ps}$.
The inset shows the the additional spectral weight in the interval
$[\tilde{\varepsilon}_U, \tilde{\varepsilon}_U+B]$. The parameters are the same
as in
fig. 1}
\label{fig2}
\end{figure}

In our calculations we used a semielliptical band density of states
$2\sqrt{B^2-\varepsilon^2}/\pi B^2$ with $B=6$. The numerical results
presented in this paper are restricted to the zero temperature limit.
Compared to the $U=\infty$ limit the spectral functions $\rho_K$
corresponding to the two time auxiliary fields provide a new input
for the calculation of the psuedo $f$-Greensfunction, which is given
by $G_{ff}$. In figure \ref{fig1} we show the spectral functions $\rho_F$,
$\rho_{\bar{F}}$ and $\rho_X$ for $N_f=14$, $\varepsilon_f=-2.5$, $U=8$
and $\tilde{\Delta}\equiv 2\tilde{V}^2/B=2.67$. The spectral functions are zero
for
energies less then $\tilde{\varepsilon}_U\equiv 2\varepsilon_f+\lambda+U\approx
6.25$ which
is determined by $\lambda_{SP}\approx 3.25$. In the following the subscripts
$SP$ will be dropped. These spectral functions together with $\lambda$
and $b$ determine the pseudo $f$-spectral function $(\rho_{ff})_{ps}
\equiv \rho_{ff}$ defined in (\ref{3.25}) which is shown in fig
\ref{fig2}.
The main spectral feature is the "Kondo - peak" slightly above the
chemical potential taken as the zero of energy. As can already be
seen in the limit $N_f\rightarrow\infty$ (\ref{3.16}) and discussed in
detail in \cite{13} the peak position ist \underline{not} given by
$\varepsilon_f+\lambda$ as (for a flat band density of states) in the
$U=\infty$ limit. There is additional spectral weight in the interval
$[\tilde{\varepsilon}_U, \tilde{\varepsilon}_U+B]$ which is magnified
in the inset of the figure. This is
$f^2$ - weight is very \underline{small} for the \underline{pseudo}
$f$ - spectral function, while it is \underline{large} for the real
$f$ - spectral function as discussed in references \cite{13} and
\cite{14}. The real $f$ - spectral function $(\rho_{ff})_{real}$ in
the saddle point approximation consists of one contribution
$|b|^2(\rho_{ff}(\varepsilon))_{ps}$, i.e. the Kondo peak position in
$(\rho_{ff})_{real}$ is determined by the peak position in
$(\rho_{ff})_{ps}$. There are additionals terms in the expression for
the real $f$ - Greens function \cite{14} which are responsible for
the large $f^2$ - weight in $(\rho_{ff})_{real}$.

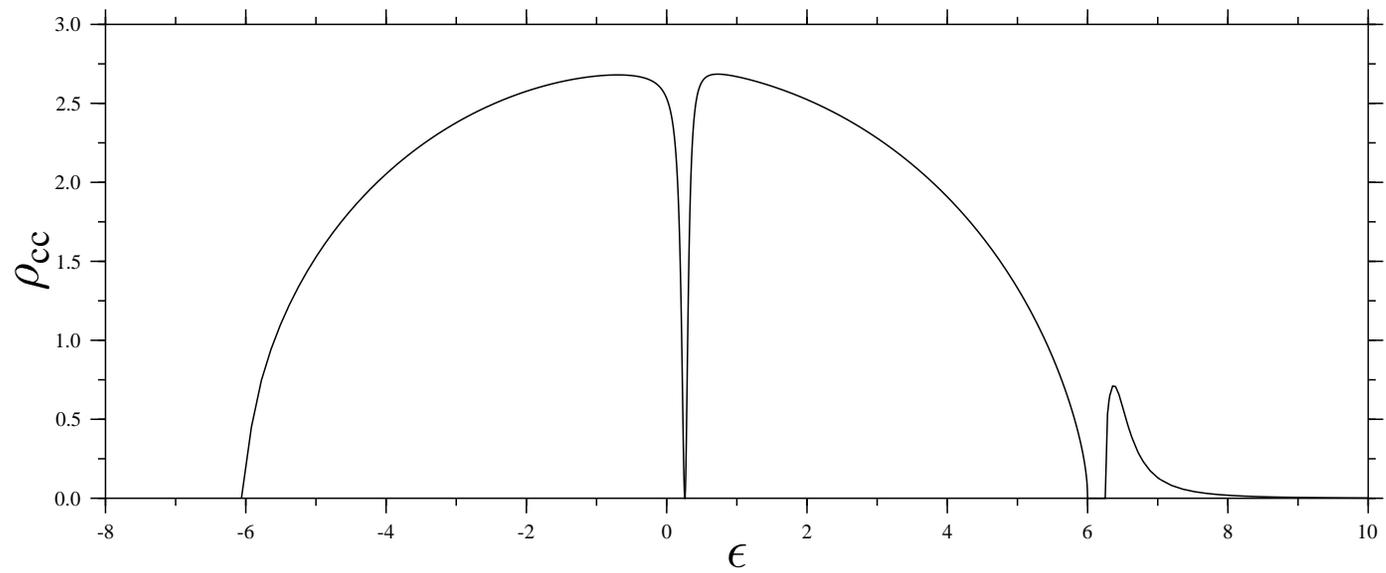
\begin{figure}
\caption{$\rho_{cc}(\varepsilon)$ and $\rho_{fc}(\varepsilon)$ are shown.
The parameters are the same as in fig. 1}
\label{fig3}
\end{figure}

The spectral functions
$\rho_{cc}(\varepsilon)$ and $\rho_{fc}(\varepsilon)$ are shown in fig
\ref{fig3}.
This dip in $\rho_{cc}(\varepsilon)$ at the position of the Kondo peak
is readily understood from Eqn. (\ref{3.13}). The occupancies
$\langle P_n \rangle$ for the parameter values used in fig.
\ref{fig1}-\ref{fig3} are
$\langle P_0 \rangle=0.144$, $\langle P_1 \rangle=0.792$, $\langle P_2
\rangle=0.063$. This corresponds to the real $f$ - occupancy
$(n_f)_{real} = \langle P_1 \rangle +2\langle P_2 \rangle=0.92$.

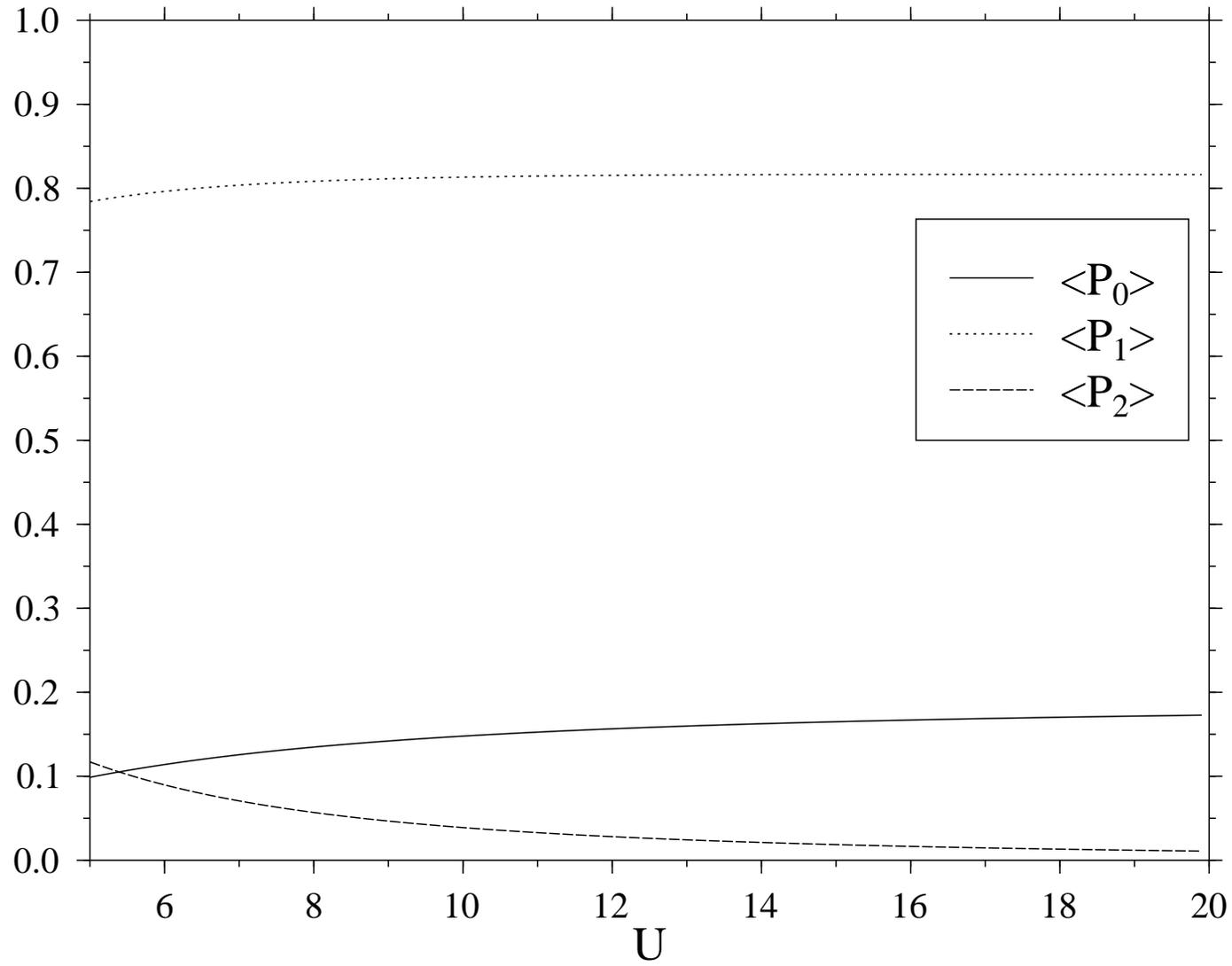
\begin{figure}
\caption{The occupation probabilities $\langle
P_n \rangle$ for $N_f=2$, $\varepsilon_f=-2.5$ and $\tilde{\Delta}=2.67$
as a function of $U$.}
\label{fig4}
\end{figure}

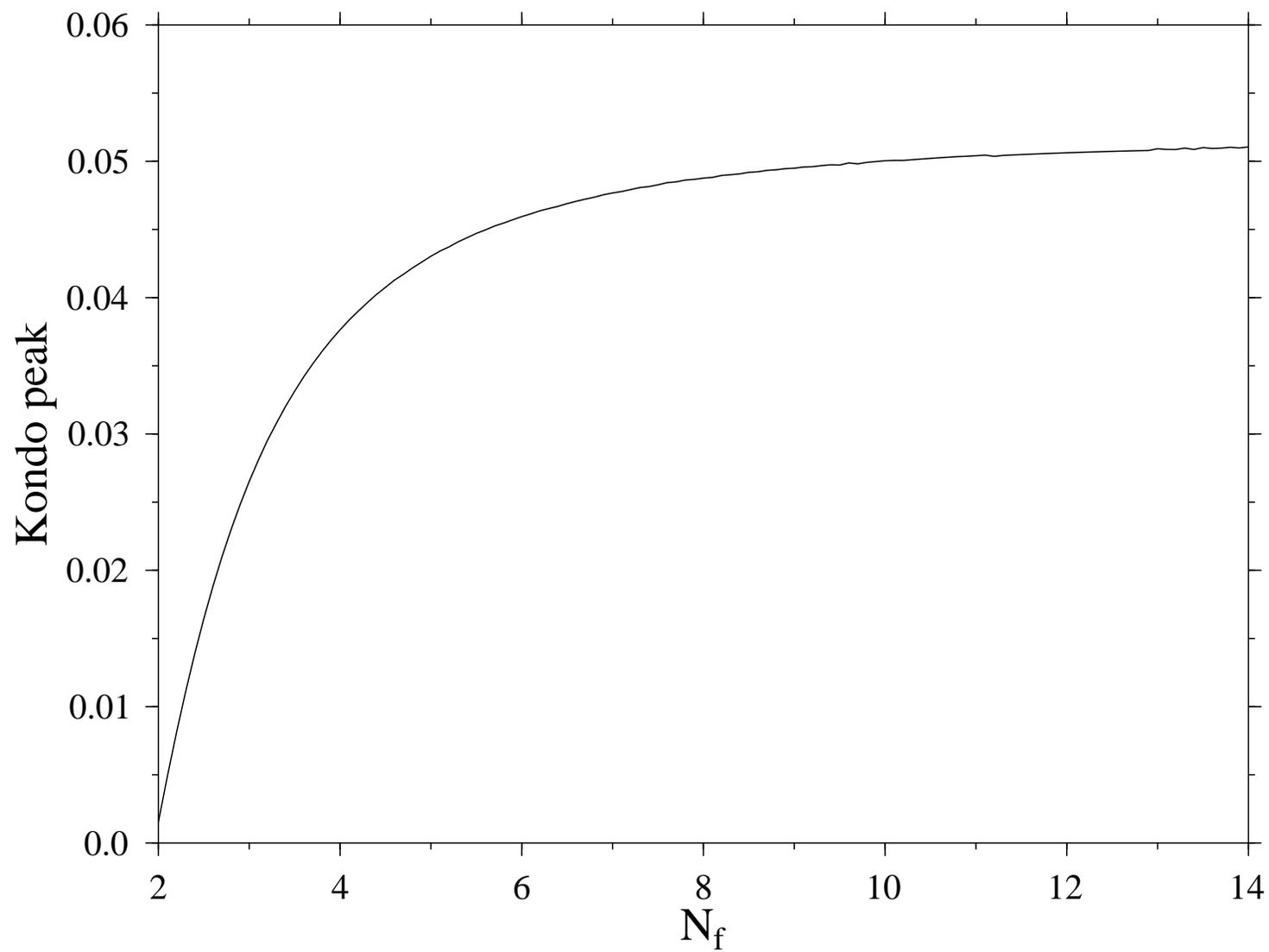
\begin{figure}
\caption{The position of the Kondo peak as a function of $N_f$ for
$\varepsilon_f=-2.5$, $U=5$ and $\tilde{\Delta}=1.33$.}
\label{fig5}
\end{figure}

The results shown so far correspond to $N_f=14$, i.e. the
\underline{large} $f$ - degeneracy obtained by neglecting spin -
orbit and crystal field splitting. As there are no exact results
available for the model for arbitrary $N_f$ and $U$ the deviation
of the the SPE results from the exact ones cannot be judged.
A very stringent test of the quality of the saddle point results can be
made for the most interesting value $N_f=2$ of the degeneracy, i.e.
the spindegenerate model, where exact results are available as
discussed in the introduction. As the SPE provide the exact solution
in the limit of \underline{infinite} degeneracy one can hardly expect
reliable results down to $N_f=2$. The exact solution for $N_f=2$ in
the symmetric case $2\varepsilon_f+U=0$ and a symmetric half-filled band
has particle-hole symmetry, which leads to $\langle P_0 \rangle
=\langle P_2 \rangle$ and a symmetric real $f$ - spectral function,
i.e. the Kondo peak is located at the energy zero. In our approach
the $f^0$ and the $f^2$ subspaces are treated in completely
\underline{different} ways and we there \underline{cannot} expect the
solution of the SPE for the symmetric $N_f=2$ case to obey particle -
hole symmetry. To test how well the behavior from the exact solution
is reproduced
by the solution of the SPE we show in fig. \ref{fig4} the probabilities
$\langle
P_n \rangle$ for $N_f=2$, $\varepsilon_f=-2.5$ and $\tilde{\Delta}=2.67$
as a function of $U$ for $U>5$. In the exact solution the crossing
between  $\langle P_0 \rangle$ and $\langle P_2 \rangle$ curves
accurs at $U=5$, while the SPE crossing ist at $U=5.4$. If one has
symmetric parameters $2\varepsilon_f+U$ but $N_f>2$ the exact Kondo peak
lies at a small positive energy and its position for
$N_f\rightarrow\infty$ is given by the solution of the SPE. While the
peak position of the exact solution is only defined for integer
values of $N_f$ we can study the peak position from the SPE as a
continuous function of $N_f$. This is shown in fig. \ref{fig5} for
$\varepsilon_f=-2.5$, $U=5$ and $\tilde{\Delta}=1.33$. With the
degeneracy approaching $N_f=2$ the Kondo peak approaches zero, but
does not quite reach it.

The results shown in figs. \ref{fig4} and \ref{fig5} indicate that the SPE are
quite
good even for rather small values of the degeneracy. A more detailed
presentation of results for thermodynamic properties and the real $f$
- spectral functions also at finite temperatures will be presented in
a forthcoming publication.

\section{Summary}

The generalized mean field theory for the Anderson impurity model
including double occupancy presented in this paper has filled a well
known gap in the approximate treatments of this model. In contrast to
the mean field theory for the $U=\infty$ model our new approximation
cannot alternatively be obtained from a simple factorization of
higher correlations. The use of the coherent state functional
integral and the introduction of two time auxiliary fields seems
necessary to obtain the correct structure in the large
degeneracy limit. In the saddle point approximation the theory yields
the exact result for the ground state in the
limit $N_f\rightarrow\infty$. For large but finite $N_f$ the solution
of the SPE is expected to provide a good approximation for
temperatures small compared to the Kondo temperature.
The spindegenerate limit cannot be reproduced exactly,
as the theory treats the $f^0$ - and$f^2$ - subspaces in completely
different ways. The presented mean field theory can be generalized in
a rather straightforward way from the impurity to the lattice model.

\begin{appendix}

\section{}

In this appendix we present a short but selfcontained derivation of
the integral equation which has to be solved to obtain the exact
ground state in the infinite degeneracy limit \cite{13}. This equation also
describes the exact ground state for {\bf finite} $N_f$ if the
valence band is {\bf completely filled}.

Let $\mid 0>$ be the state with all conduction states below the Fermi
energy, $\varepsilon_F = 0$, filled and the $f$-level empty. This
state couples via $H_A$ to the states

\begin{equation} \label{a1}
\mid \varepsilon > \equiv \frac{1}{\sqrt N_f} \sum_\nu f^+_\nu
c_{\varepsilon\nu}\mid 0 >,
\end {equation}

in which one conduction electron below $\varepsilon_F$ has hopped
into the $f$-level. These states couple to states in which the
$f$-level is doubly occupied

\begin{equation} \label{a2}
\mid \varepsilon, \varepsilon'> =
\frac{1}{\sqrt {N_f(N_f-1)}} \sum_{{\nu,\nu'}\atop{\nu\neq \nu'}}
f^+_\nu c_{\varepsilon\nu}f^+_{\nu'} c_{\varepsilon'\nu'}
\mid 0>,
\end{equation}

as well as to states with the $f$-level empty and a particle
hole-pair in the conduction band. The coupling to the latter states
is smaller by a factor $1/\sqrt{N_f}$ and can therefore be neglected
in the large $N_f$ limit \cite{13}. In this limit the ground state takes the
form

\widetext
\begin{equation} \label{a3}
\mid \phi_0> = A \left[ \mid 0> + \int^0_{-B} d \varepsilon \, a(\varepsilon)
\mid \varepsilon > + \int^0_{-B} d\varepsilon \int^\varepsilon_{-B}
d\varepsilon' b(\varepsilon, \varepsilon') \mid
\varepsilon,\varepsilon'> \right]
\end{equation}
\narrowtext

where $A$ is the normalization constant. Inserting into the
Schr\"odinger equation $H_A\mid\phi_0 > = E\mid\phi_0>$ the
$b(\varepsilon,\varepsilon')$ can be simply expressed in terms of
$a(\varepsilon)$ and $a(\varepsilon')$ and one obtains the equations
\cite{13}

\begin{equation} \label{a4}
\Delta E = \int^0_{-B} \tilde{V}(\varepsilon)
a(\varepsilon)d\varepsilon,
\end{equation}

\widetext
\begin{equation} \label{a5}
\left[\varepsilon_f - \varepsilon - \Delta E -
(1 - \frac{1}{N_f}) \tilde \Gamma (2\varepsilon_f + U - \varepsilon -
\Delta E) \right]a(\varepsilon)
+ \tilde V (\varepsilon)
\left[1- (1 - \frac{1}{N_f})\int^0_{-B}
\frac{\tilde V(\varepsilon_1) a(\varepsilon_1)}
{2\varepsilon_f + U - \varepsilon - \Delta E - \varepsilon_1}
d\varepsilon_1 \right] = 0
\end{equation}
\narrowtext

where $\Delta E \equiv E_0 - <0\mid H\mid 0>, \tilde V(\varepsilon) =
\sqrt {N_f} V(\varepsilon)$ and

\begin{equation} \label{a6}
\tilde \Gamma (z) = \int ^0_{-B} \frac{\tilde V (\varepsilon)^2}
{z - \varepsilon}
\end{equation}

In the derivation the matrix elements $<\varepsilon\mid H\mid0> =
\tilde V(\varepsilon)$ and

\widetext
\begin{equation} \label{a7}
<\varepsilon,\varepsilon' \mid H \mid \varepsilon''> =
(1-\frac{1}{N_f}) \left[\tilde V (\varepsilon') \delta (\varepsilon
- \varepsilon'')
+ \tilde V(\varepsilon) \delta (\varepsilon' - \varepsilon'')\right]
\end{equation}
\narrowtext

are used. For the comparison with the saddle point equations we
define the function $c(\varepsilon)$ for $-B < \varepsilon < 0$ by

\widetext
\begin{equation} \label{a8}
a(\varepsilon) \equiv \tilde V(\varepsilon)
\frac{1+c(\varepsilon)}
{\Delta E - \varepsilon_f + \varepsilon + (1-\frac{1}{N_f})\tilde
\Gamma (2\varepsilon_f + U - \varepsilon - \Delta E)}
\end{equation}
\narrowtext
Then equations (\ref{a4}) and (\ref{a5}) read

\widetext
\begin{equation} \label{a9}
\Delta E = \int^0_{-B} \frac{\mid \tilde V (\varepsilon)\mid^2
(1 + c (\varepsilon))}
{\Delta E - \varepsilon_f +\varepsilon + (1-\frac{1}{N_f})\tilde
\Gamma (2\varepsilon_f + U - \varepsilon - \Delta E)}
\end{equation}
\begin{equation} \label{a10}
c(\varepsilon) + (1 - \frac{1}{N_f}) \int^0_{-B}
\frac{1}{(2\varepsilon_f + U - \varepsilon - \Delta E - \varepsilon_1)}
\frac{\mid \tilde V (\varepsilon_1) \mid^2 (1+c(\varepsilon_1))}
{\Delta E - \varepsilon_f + \varepsilon_1 + \tilde \Gamma
(2\varepsilon_f + U - \varepsilon_1 - \Delta E)} d\varepsilon_1 = 0
\end{equation}
\narrowtext

In the limit $\varepsilon_f + U >> B$ the integral equation
(\ref{a10}) is
of separable form. The function $c(\varepsilon)$ for
$-B<\varepsilon<0$ can be analytically continued using Eqn.
(\ref{a10}).
Then $c(z)$ has a branch cut on the real axis from
$2\varepsilon_f + U - \Delta E$ to $2\varepsilon_f + U - \Delta E + B$. The
saddle point equations reduce at $T = 0$ in the large degeneracy
limit to Eqn. (\ref{a9}) and (\ref{a10}) if one identifies $c(\varepsilon)$
with $\bar x (\varepsilon)$ and $\Delta E$ with $-\lambda$.

\section{}

In this appendix we derive the expression for the grand canonical
potential at the saddle point $J_{SP}$ and discuss
how it simplifies in the limit $N_f \rightarrow \nolinebreak \infty$.

We begin with the evaluation of the last term on the rhs of Eqn.
(\ref{2.17}) for the effective action $S_{eff}$. At the saddle point the
fields $F,\bar F, X \mbox{ and } \bar X$ depend on the time difference
only and the "time-part" of the trace can be written as a sum over
the fermionic Matsubara frequencies. If we denote the unperturbed
propagators in accordance with Eq. (\ref{3.10}) as $G^{00}(i\omega_n)$ we
have to evaluate $tr \ln [1-G^{00}h_1]$ where the "small trace" runs
over the one electron Hilbert space. We use $tr\ln \hat a = \ln \det
\hat a$ as the determinant is easily calculated using the special
form of $h_1$ (\ref{2.18b})

\widetext
\begin{eqnarray} \label{b1}
\det(1-G^{00}(i\omega_n)h_1(i\omega_n))
& = &
(1-<f\mid G^{00}\mid f> V\bar F)
(1-<c\mid G^{00}\mid c> VF)\nonumber \\
& - &
V^2(\bar b + X)<c\mid G^{00}\mid c>(b + \bar X) <f\mid G^{00}\mid
f> \vspace{0.5cm}\nonumber \\
& = &
\frac{1-VF(i\omega_n)G^{00}_{cc}(i\omega_n)}
{(i\omega_n - \tilde \varepsilon_f) G_{ff}(i\omega_n)}
\end{eqnarray}
\narrowtext

In the second equality we have used (\ref{3.11}) and (\ref{3.12a}). The
Matsubara
sum is now performed as a contour integral using the assumptions
about the analytic continuation of the auxiliary fields discussed in
section II. This yields

\widetext
\begin{equation} \label{b2}
-\frac{1}{\beta}\cdot N_f Tr\ln [\partial \otimes 1_e + 1_\tau
\otimes h_0 + h_1]\mid_{SP} = J^0_M + J^0_f + (\Delta J)_1
\end{equation}
\narrowtext

with

\widetext
\begin{equation} \label{b3}
(\Delta J)_1  =  N_f \int \frac{d\varepsilon}{2 \pi i} f(\varepsilon)
\bigg\{   \ln \left[ \frac{1-VF(\varepsilon + io)G^{00}_{cc}
(\varepsilon + i0)}
{1-VF(\varepsilon - io)G^{00}_{cc}(\varepsilon - i0)}
\right]
+ \ln \left[ \frac{(\varepsilon - \tilde \varepsilon_f - io) G_{ff}
(\varepsilon - io)}
{\varepsilon - \tilde \varepsilon_f + io) G_{ff}
(\varepsilon + io)}
\right]\quad \bigg\}
\end{equation}
\narrowtext

and $J^0_M + J^0_f$ the grand canonical potential for a system
noninteracting electrons described by $h_0$.

If we denote the double integral on the rhs of (\ref{2.17}) by $\beta I$ we
obtain using the SPE

\widetext
\begin{equation} \label{b4}
I = N_f V \int \frac{d\varepsilon}{2\pi i} f(\varepsilon)
[(FG_{cc} + XG_{fc})_{\varepsilon -i0} -
(FG_{cc} + XG_{fc})_{\varepsilon +i0}]
\end{equation}
\narrowtext

Using (\ref{2.16}) with $Z^h_0(\lambda) \approx 1$ for the interesting
temperature regime we finally obtain

\begin{equation} \label{b5}
J_{SPA} = J^0_M +J^0_f + (\Delta J)_1 - I + \lambda \mid b \mid^2 -
\lambda
\end{equation}

For $T \rightarrow 0$ this expression simplifies considerably in the
large $N_f$ limit. Then $VF$ and $G_{ff}- G^0_{ff}$ are of order
$1/N_f$ and the logarithns in (\ref{b3}) can be expanded. To the same
order in $1/N_f$ the Greens functions in (\ref{b4}) can be replaced by the
unperturbed functions $G^{00}_{cc}$ and $G^0_{fc}$. Then in order
$(1/N_f)^0$ all terms in the expression for $J_{SPA}$ cancel except
the first and the last one. With $E^{(0)}_0 \equiv <0\mid H\mid 0>$
this yields

\begin{equation} \label{b6}
(E_0)^{SPA} = E^{(0)}_0 -\lambda
\end{equation}

as mentioned following Eqn. (\ref{3.24a}-\ref{3.24last}).

\section{}

In this appendix we discuss how the "real $f$-occupancy" in the
saddle point approximation $(n_f)^{SPA}_{real}$ can be
expressed in terms of the "pseudo-$f$-occupancy"
$(n_f)^{SPA}_{ps} = N_f G^{SPA}_{ff}(\tau = -0)$. The
calculation shows that the constant $A$ defined following Eq.
(\ref{3.5})
presents the probability for the double occupancy of the $f$-level.

The real $f$-occupancy can be calculated from the grand canonical
potential $J = -\ln Z/\beta$ with $Z$ defined in (\ref{2.6}) by
differentiation with respect to the $f$-level energy $\varepsilon_f$

\begin{equation} \label{C.1}
(n_f)_{real} = \frac{\partial J}{\partial\varepsilon_f}
\end{equation}

In the $SPA$ the grand canonical potential is given in (\ref{b5}). With $h
\equiv h_0 + h_1$ it reads

\begin{eqnarray} \label{C.2}
J_{SPA} & = & -(N_f/\beta)Tr\ln(\partial+h)-I+\lambda
(\mid b \mid^2-1) \nonumber \\
& \equiv & J_h -I+\lambda (\mid b \mid^2-1)
\end{eqnarray}

In the following the label SPA at all fields and Greens functions
will be suppressed. The contribution to $(n_f)_{real}$ from the first
term on the rhs of (C.2) is
\begin{equation} \label{C.3}
\frac{\partial J_h}{\partial \varepsilon_f} =
(N_f/\beta) Tr(G\frac{\partial h}{\partial \varepsilon_f})
\end{equation}

and with $h = 1_\tau \otimes h_0 + h_1$ given in
(\ref{2.18a}-\ref{2.18last}) we obtain

\widetext
\begin{eqnarray} \label{C.4}
\frac{\partial J_h}{\partial \varepsilon_f} & = &
(1+\frac{\partial \lambda}{\partial \varepsilon_f})(n_f)_{ps}
+\frac{1}{\beta}\big[
N_f V tr_{\tau}(G_{ff}\frac{\partial \bar F}{\partial \varepsilon_f})
+N_f V tr_{\tau}(G_{cc}\frac{\partial F}{\partial \varepsilon_f})
\nonumber \\
& + & N_fV\big(\frac{\partial \bar b}{\partial \varepsilon_f}G_{fc}(0)
+\frac{\partial b}{\partial \varepsilon_f}G_{cf}(0)\big)
+N_fV tr_{\tau}\big(G_{fc}\frac{\partial X}{\partial \varepsilon_f}+
G_{cf}\frac{\partial \bar X}{\partial \varepsilon_f}\big)\big]
\end{eqnarray}
\narrowtext

where $tr_\tau$ denotes the "time"-integration part of the trace.
Differentiating the double integral (\ref{b4}) yields

\widetext
\begin{eqnarray} \label{C.5}
\frac{\partial I}{\partial \varepsilon_f} & = & \frac{1}{\beta}
\int d\tau d\tau'\frac{1}{G^0_h(\tau-\tau')}
\big[\frac{\partial\bar X(\tau-\tau')}{\partial \varepsilon_f}X(\tau-\tau')
+\bar X(\tau-\tau')\frac{\partial X(\tau-\tau')}{\partial \varepsilon_f}
\nonumber \\& & +
\frac{\partial \bar F(\tau-\tau')}{\partial \varepsilon_f}F(\tau-\tau')
+\bar F(\tau-\tau')\frac{\partial F(\tau-\tau')}{\partial \varepsilon_f}
\big] -(2+\frac{\partial \lambda}{\partial \varepsilon_f}) A
\end{eqnarray}
\narrowtext

and from the last term on the rhs of (C.2) we obtain

\begin{equation} \label{C.6}
\frac{\partial}{\partial \varepsilon_f}\lambda(\mid b\mid^2-1) =
\frac{\partial \lambda}{\partial \varepsilon_f}(\mid b\mid^2-1)
+\lambda(\frac{\partial \bar b}{\partial \varepsilon_f}b+
\bar b \frac{\partial b}{\partial \varepsilon_f})
\end{equation}

If we add up all contributions and use the SPE (\ref{3.3a}-\ref{3.3last}) and
(\ref{3.5}) we
obtain the simple result

\begin{equation} \label{C.7}
(n_f)_{real}=(n_f)_{ps}+2A
\end{equation}

This suggests that $A$ is the probablity for the double occupancy of
the $f$-level. This is readily verified by calculating $<P_2>$,
where $P_n$ is projection operator for the $f$-level occupied by $n$
electrons. A calculation very similar to the one presented above
yields in SPA

\begin{equation} \label{C.8}
<P_2>=\frac{\partial J}{\partial U}=A
\end{equation}

In the limit $N_f \rightarrow \infty$ and $T \rightarrow 0$ this
identity can be explicitely checked using the results of appendix A
and the expression for $<P_2>$ presented in ref. 13. From (C.7)
and (C.8) and $\sum^\alpha _{n=0} < P_n> = 1$ we also obtain $<P_1> =
(n_f)_{ps}$ and $<P_0> = \mid b \mid^2$, which elucidates the meaning
of the SPE (\ref{3.5}).

The generalization of (C.8) beyond the SPA is given by using
(\ref{2.16})
to evaluate $\partial J/\partial U$. This yields

\widetext
\begin{equation} \label{C.9}
<P_2>=\frac{1}{Z}\int_C \frac{\beta d\lambda}{2\pi i}e^{\beta \lambda}
\int Db\, DF\, DX  \nonumber \\
\big( \int \int d\tau d\tau'
\frac{\mid X(\tau,\tau') \mid^2 +\mid F(\tau,\tau') \mid^2}
{G^0_h(\tau-\tau')}\frac{\partial}{\partial \lambda}
G^0_h(\tau-\tau')\big) e^{-S_{eff}}
\end{equation}
\narrowtext

where we have taken $Z^h_0(\lambda)\approx 1$ as usual. If we take
the $FI$ on the rhs of (C.9) as the generalized expression for $A$,
equation (C.7) also holds generally.

\end{appendix}


The spectral functions

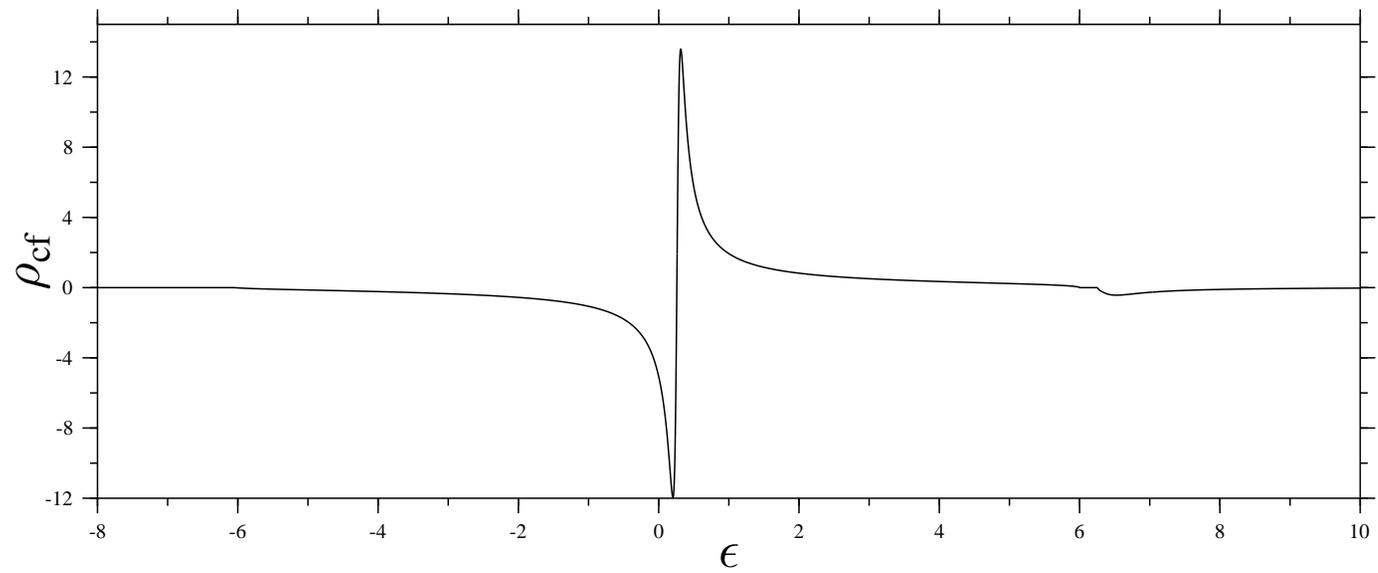
\end{document}